\documentclass[epjST]{svjour}
\usepackage{graphics}
\usepackage[usenames]{color}
\usepackage{textcomp} % \textmu
\usepackage{xspace} % \textmu
\newcommand{\um}{{\textmu}m\xspace}

\begin{document}
\title{Family-Viscek scaling of detachment fronts in Granular Rayleigh-Taylor instabilities during sedimentating granular/fluid flows}

\author{Jan Ludvig Vinningland\inst{1,2} \and Renaud Toussaint\inst{3} \fnmsep\thanks{\email{renaud.toussaint@unistra.fr}}\and Michael Niebling \inst{1,3} \and Eirik G. Flekk{\o}y \inst{1} \and Knut J{\o}rgen M{\aa}l{\o}y \inst{1}}
%\fnmsep\thanks{\email{j.l.vinningland@fys.uio.no}}
\institute{Department of Physics, University of Oslo, P.O. Box 1048, 0316 Oslo, Norway \and International Research Institute of Stavanger, IRIS, P.O. Box 8046, 4068 Stavanger, Norway \and IPGS, University of Strasbourg, CNRS, 5 rue Descartes, 67000 Strasbourg}
\abstract{ When submillimetric particles saturated with a fluid form a
  compact cluster on the top on a confined clear fluid of the same
  nature, they fall by detaching from the lower boundary of the
  cluster, and this separation front between particles and fluid is
  unstable.  Particles detach and fall in the clear fluid region,
  giving rise to growing fingers of falling particles.  We study this
  problem using both experiments and hybrid granular/fluid mechanics
  models.  In the case of particles from 50 to 500 microns in diameter
  falling in air, we study the horizontal density fluctuations at
  early times: the amplitude of the density difference between two
  points at a certain horizontal distance grows as a power law of
  time. This happens up to a saturation corresponding to a power law
  of the distance. The way in which the correlation length builds up to
  this saturation also follows a power law of time. We show that these
  decompaction fronts in sedimentation problems follow a Family-Vicsek
  scaling, characterize the dynamic and Hurst exponent of the lateral
  density fluctuations, respectively $z \sim 1$ and $\zeta \sim 0.75$,
  and show how the prefactors depend on the grain diameter.  We also
  show from similar simulations with a more viscous and incompressible
  fluid, that this feature is independent of the fluid compressibility
  or viscosity, ranging from air to water/glycerol mixtures.
} %end of abstract
\maketitle
\section{Introduction}
\label{intro}
In many natural and industrial situations, mixtures of granular
materials and fluids are deformed.  Physicists have intensively
studied granular mechanics using discrete methods over the last 30
years \cite{Herrmann,Duran} since the rise of powerful computers,
with various types of avalanche dynamics \cite{altshuler08},
the dynamics of sheared layers \cite{aharonov02,aharonov04} and the
stress distribution in sandpiles \cite{Clement}. For sufficiently
small grains, when the fluid seeps through the granular assembly, the
resulting drag is sufficient to rearrange the grains, and it is
necessary to take into account the momentum exchange between the
granular and the fluid phase to understand the dynamics.

Such hybrid granular/fluid flows are important/manifest in many
natural flows at the surface of the earth: e.g. in liquefaction
\cite{goren10}, mudflows \cite{travelletti11,malet,vanasch}, or sand
volcanoes. They also play a role in underwater avalanches and
turbidites, or in sedimentation processes when large amounts of
sediments are released and deposited. Eventually, their understanding
is/are crucial for the mechanical stability of/in borehole
exploitation/exploration, during pumping in to or out of man-made
wells, in hydrofracture \cite{flekkoy02}, or for the understanding
of reservoir stability and fault lubrication \cite{goren11}.

Elementary situations of sedimenting grains/unconsolidated grains (in
a fluid) and their stability have been recently studied in generic
situations where a dense granular packing saturated with a fluid is
subjected to a flow of the same fluid.  For example, the propagation
of rising bubbles in air-grain flows in tubes, or the
intermittency of flows in ticking hourglasses has been studied
\cite{LePennec,flekkoy01,gendron01}.  Experiments with dense granular packings
displaced by pressurized air have been followed optically
in Hele-Shaw cells
\cite{johnsen06,johnsen07,johnsen08,vinningland07a,niebling10a}, and
studied numerically using models where the momentum exchange is based on a
Darcy description, with a flow described at a coarse grained scale of
a few grain diameters \cite{mcnamara00,anghel06}.

For example, when a granular packing in a circular cell where gravity
acts perpendicularly to the cell walls starts to flow due to a
sufficient air-overpressure in the centre, an instability
corresponding to a granular Saffman-Taylor analog was observed
\cite{johnsen06,cheng07}. Such analog behavior was also observed in
linear cells, either with a free boundary at the outlet leading to
decompaction \cite{johnsen07,johnsen08}, or with a confined flow for
the particles leading to compaction
\cite{niebling11,schelstraete09}. Changing the properties of the
interstitial fluid with orders of magnitudes in viscosity and
compressibility (between oil and air) was shown to affect the dynamics
of this process, but to lead to the formation of similar patterns
\cite{johnsen08}.

When gravity is acting along the flow, a granular analog of the
Rayleigh-Taylor instability has been studied: situations where
initially a pack of dense grains are released into a zone of lighter
fluid. The same fluid is also present in the interstitial space
between the packed grains. This situation would naturally arise during
sedimentation flows, e.g. when a sudden amount of compact material is
released in open water. This instability has been studied with
mixtures of grains (typically 100 \um in diameter) and
air \cite{vinningland07a,vinningland07b,vinningland10}, and with less
compressible and more viscous fluids such as water glycerol or oil
\cite{niebling10a,niebling10b,voltz01}.

The release of such dense packs of grains in a clear fluid contained
in a closed cell, leads to the formation of a detachment front at the
bottom of the pack. The particles fall faster from the bottom part
than from the upper part of the pack, and fingers form in the
previously clear fluid region. This front is unstable, grows fast in
amplitude, and leads to coarsening bubbles empty of grains rising
through the grain pack.  The dynamics of this front has been studied
using Fourier analysis by investigating the growth of each mode
\cite{vinningland07a,vinningland07b,vinningland09a,vinningland09b}. In this system, the change of
the particle size, for sizes ranging from roughly 50 to 500 $\mu$m in
diameter, was shown to leave the system dynamics invariant, using the
grain diameter as the characteristic length scale.

We will show in this manuscript that the horizontal density profiles,
as function of time and space, for different particle sizes, can be
entirely collapsed and follow a Family-Vicsek scaling. Such scaling,
introduced by Family and Vicsek \cite{family85}, was found to be
obeyed by many surface growth processes. It relates the anomalous
diffusion of a quantity, the growth of correlations, and the self
affine geometry of the developed interface - or of the quantity
characterized by this scaling \cite{Barabasi}. For example, such a collapse was
observed for the thermal fluctuations of magnetic hole assemblies,
which are dipolar chains in thin fluid cells
\cite{toussaint04a,toussaint04b,toussaint06}. It is also obeyed by
natural patterns slowly growing in rocks, as pressure solution seams
evolving into stylolites
\cite{schmittbuhl04,koehn2007,rolland11,laronne}.

This implies that a correlation length grows in this system, as a power
law of time, $\xi\sim t^{1/z}$, with a dynamic exponent around $z\sim
1$. For scales above $\xi$, the root mean square of the lateral
density fluctuations grows in a super diffusive way, as $\sigma \sim
t^\beta$ with a growth exponent around $\beta \sim 0.75$. Below the
correlation length, the density profiles are shown to be self-affine,
with a Hurst exponent around $\zeta \sim 0.75$.

In the next section, we will summarize the methodologies used to study
the granular Rayleigh-Taylor instability, from an experimental and
from a numerical perspective. Next, we will describe the types of
patterns obtained at different particle sizes. Eventually, we will
perform the Fourier analysis of the density fluctuations, and show
that they can be collapsed at different times, for different wave
numbers and for all investigated particle sizes on a master curve,
displaying the fact that they follow a Family-Vicsek scaling.

\begin{figure}
  \begin{center}
    \resizebox{\columnwidth}{!}{%
      \includegraphics{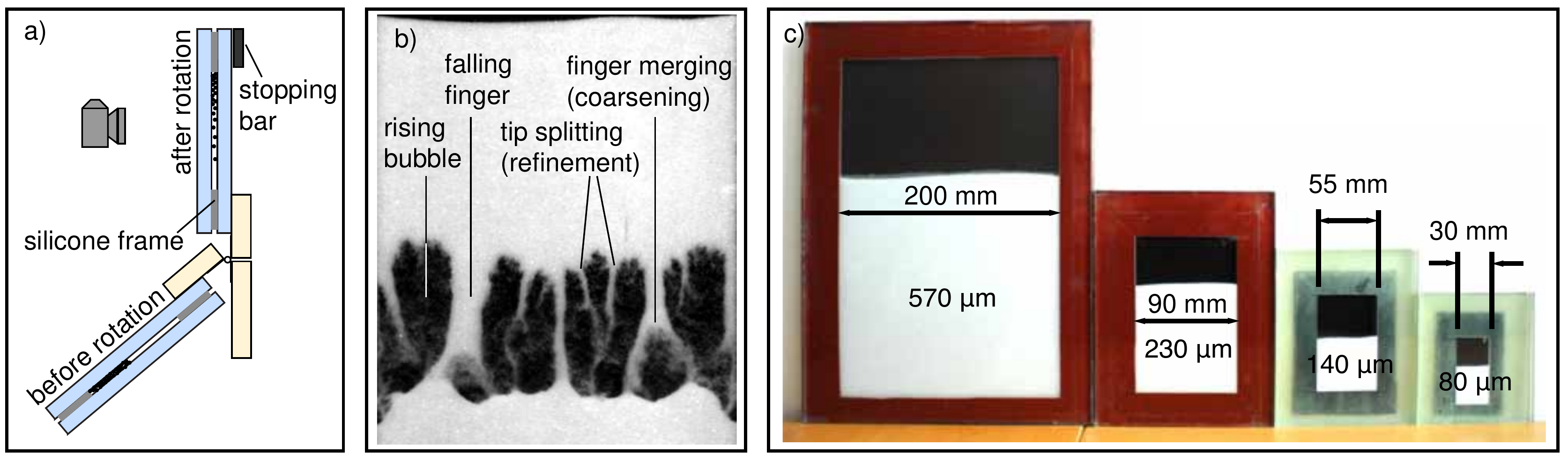} }
  \end{center}
  \caption{(a) Experimental setup and principle, (b) typical granular
    Rayleigh-Taylor flow, where fingers of falling grains (white) form
    in the previously air-filled region (black) during an experiment,
    and (c) the experimental cells of different sizes with grains of
    proportional diameters.}
  \label{Exps}       % Give a unique label
\end{figure}

\section{Rayleigh-Taylor instability: definition and methods}
\subsection{Experimental setup}
\label{Expsetup}

Historically, the Rayleigh-Taylor instability arises from a flat
interface when a dense fluid is initially situated above a lighter
one: in a closed container, the two fluids interchange positions while
forming a fingering pattern. This instability, and the selection of a
preferred wavelength due to the interplay between gravitational,
capillary and viscous forces was first studied by Lord Rayleigh and
I. G. Taylor \cite{Rayleigh83,Drazin}. The granular analog of this
instability has been studied recently with air or more viscous fluids:
it arises when initially a pack of dense grains is released on top
of clear fluid in a closed vessel. There is no surface tension in this
problem since the fluid is also initially present in the porous space
between the grains in the upper pack.

The experiments are performed in a closed impermeable Hele-Shaw cell,
which is initially prepared by letting grains accumulate at the
bottom. The cell is then suddenly rotated to be brought upside down,
and a damping stops the cell in a vertical direction, attenuating
vibrations as much as possible. The setup is illustrated in
Fig.~\ref{Exps}. Several system sizes have been used for air-grain
experiments, ranging from 80 $\mu$ m diameter beads to 570 $\mu$m
diameter ones for the air/grain system. The system sizes are
proportional to the bead diameters, from 31 to 200 mm. The grains used
in the experiments where polystyrene spheres (Ugelstad spheres from
Microbeads), with a density of 1.05 g/cm$^3$.

\begin{figure}
  \begin{center}
    \resizebox{0.75\columnwidth}{!}{%
      \includegraphics{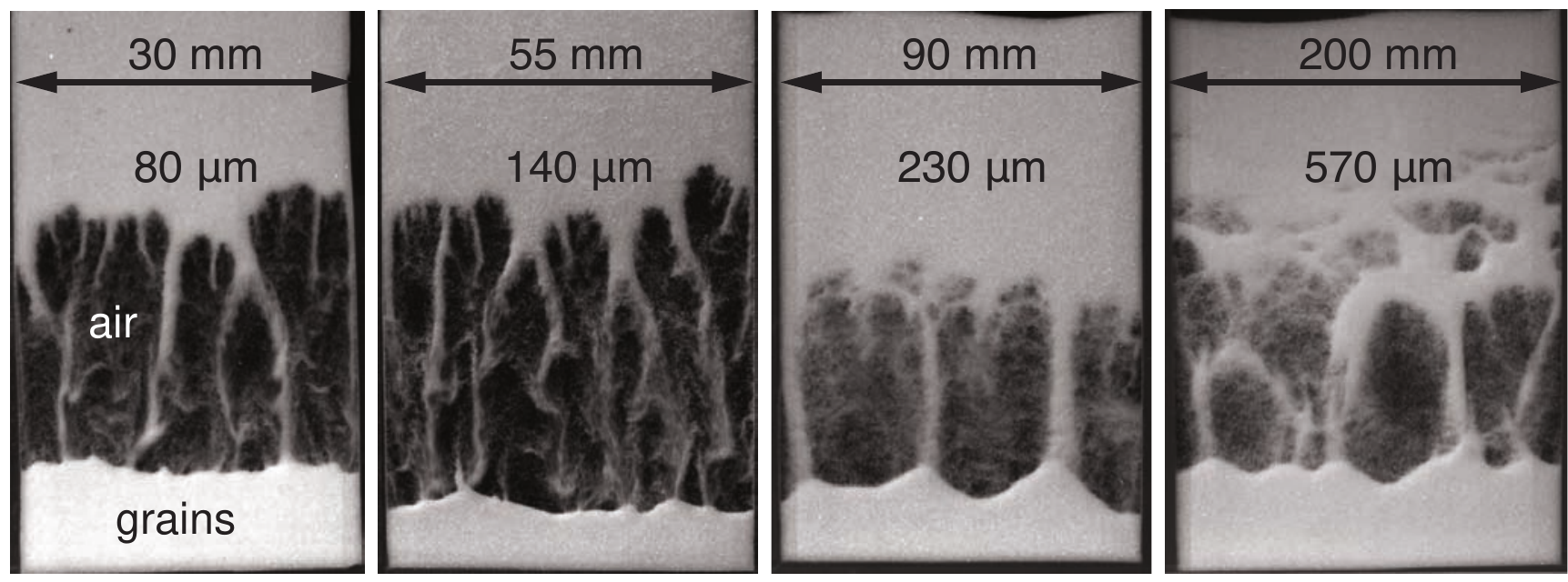} }
  \end{center}
  \caption{Experimental snapshots of the granular Rayleigh-Taylor
    instability in air where the four cells shown in
    Fig.~\ref{Exps}(c) are used.}
  \label{ExperimentalSnapshots}       % Give a unique label
\end{figure}

Early stages of the experimental pictures are shown on
Figs.~\ref{Exps} (right) and Fig.~\ref{ExperimentalSnapshots}.

\subsection{Simulation technique}
\label{Sim}

The simulation techniques are described in details in
\cite{vinningland07a} for the flow of grains and compressible air, or
\cite{niebling10a} for the flow of grains and incompressible viscous
fluid.

In summary, the principle is as follows: the grains interact via
repelling contact forces and via a drag force due to the flow of the interstitial
fluid. This can be expressed as Newton's second law for each grain, as

\begin{equation}
m a = F_I + (\rho_g-\rho_f) g - \nabla P / \rho
\end{equation}
where $m$ and $a$ are the grain mass and acceleration, $F_I$ corresponds
to the solid interaction forces with neighboring grains, the third
term corresponds to buoyancy forces, and the last one to seepage
forces: $g$ is the gravity, $\rho_f$ and $\rho_g$ are the fluid and
solid bulk mass densities, $P=P_T - \rho_f g z$ is the fluid pressure
deviating from the hydrostatic profile -- where $P_T$ is the total
fluid pressure, $\rho_f$ is the fluid density, $g$ the gravity and $z$
the depth. This pressure is evaluated over a grid discretized at the
representative elementary volume scale (the Darcy scale, a few grains
large), and $\rho$ is the number density of the grains: $\rho = V_g /
(1-\Phi)$ with $V_g$ the grain volume, and $\Phi$ the porosity.

This equation is being solved at each time step, using a Verlet
algorithm in molecular dynamic codes.  The forces $F_I$ are determined
from elastic contact model \cite{johnsen06,niebling10a}, or from a
contact dynamics scheme \cite{vinningland07a}.

The dynamic equation solved for the fluid results from the following
principles: mass conservation for the grains, mass conservation for
the fluid, equation of state for the fluid, and Darcy to get the
relative velocity between the fluid and the solid.  Calling $u$ the
grain velocity, $v$ the fluid one, and $\rho_f$ the fluid density,
this can be expressed at the Darcy scale as:
\begin{eqnarray}
\partial_t (1-\Phi) + \nabla \cdot \left(\left(1-\Phi\right) u\right) & = & 0 \\
\partial_t (\rho_f \Phi) + \nabla \cdot \left(\Phi \rho_f v\right) & = & 0 \\
v-u = - \frac{\kappa\nabla P}{\mu \Phi}
\end{eqnarray}
where $\mu$ is the fluid viscosity, $\kappa$ is the local
permeability of the packing, evaluated from the Carman-Kozeny equation:
$$\kappa = \frac{d^2}{180} \frac{\Phi^3}{(1-\Phi)^2}$$ 
and the fluid mass conservation is expressed as one of the two
alternatives: in general, linearizing this equation of state, one
gets $$\rho_f(P) = \rho_0 \left(1+ \beta \left(P-P_0\right)\right)$$ where $\beta$ is the
fluid compressibility, $P_0$ an initial (reference) pressure, and
$\rho_0$ the fluid density at that pressure.

This set of equations leads to the following reduced form:
\begin{equation}
  \Phi \left(\partial_t P + u\cdot \nabla P\right) = \nabla \cdot \left(\hat{P} \frac{\kappa}{\mu}\nabla P\right) - \hat{P} \nabla \cdot u 
\label{genpressure}
\end{equation}
where $\hat{P}=\rho_f/(\rho_0 \beta) = P - P_0 + 1/\beta$

Considering air as a perfect gas, this reduces to 
$\rho_f = \rho_0 P/P_0,$ $\beta_T = 1/P_0$ and  $\hat{P} = P,$ so that 
\begin{equation}
\Phi \left(\partial_t P + u\cdot \nabla P\right) = \nabla \cdot \left(P \frac{\kappa}{\mu}\nabla P\right) - P \nabla \cdot u
\label{airpressure}
\end{equation}
which is solved alternately with the molecular dynamic time steps,
using a Cranck-Nicholson algorithm \cite{Press}.

For a viscous fluid considered as incompressible, we have $\beta=0$,
$\rho_f = \rho_0. $ and the equation \ref{genpressure} in the limit
$\hat{P}\rightarrow \infty$, reduces to
$$ \nabla \cdot \left(\frac{\kappa}{\mu}\nabla P\right) =  \nabla \cdot u $$
which is solved at each time step using a multigrid algorithm \cite{Press}.

\begin{figure}
  \begin{center}
    \resizebox{.9\columnwidth}{!}{%
      \includegraphics{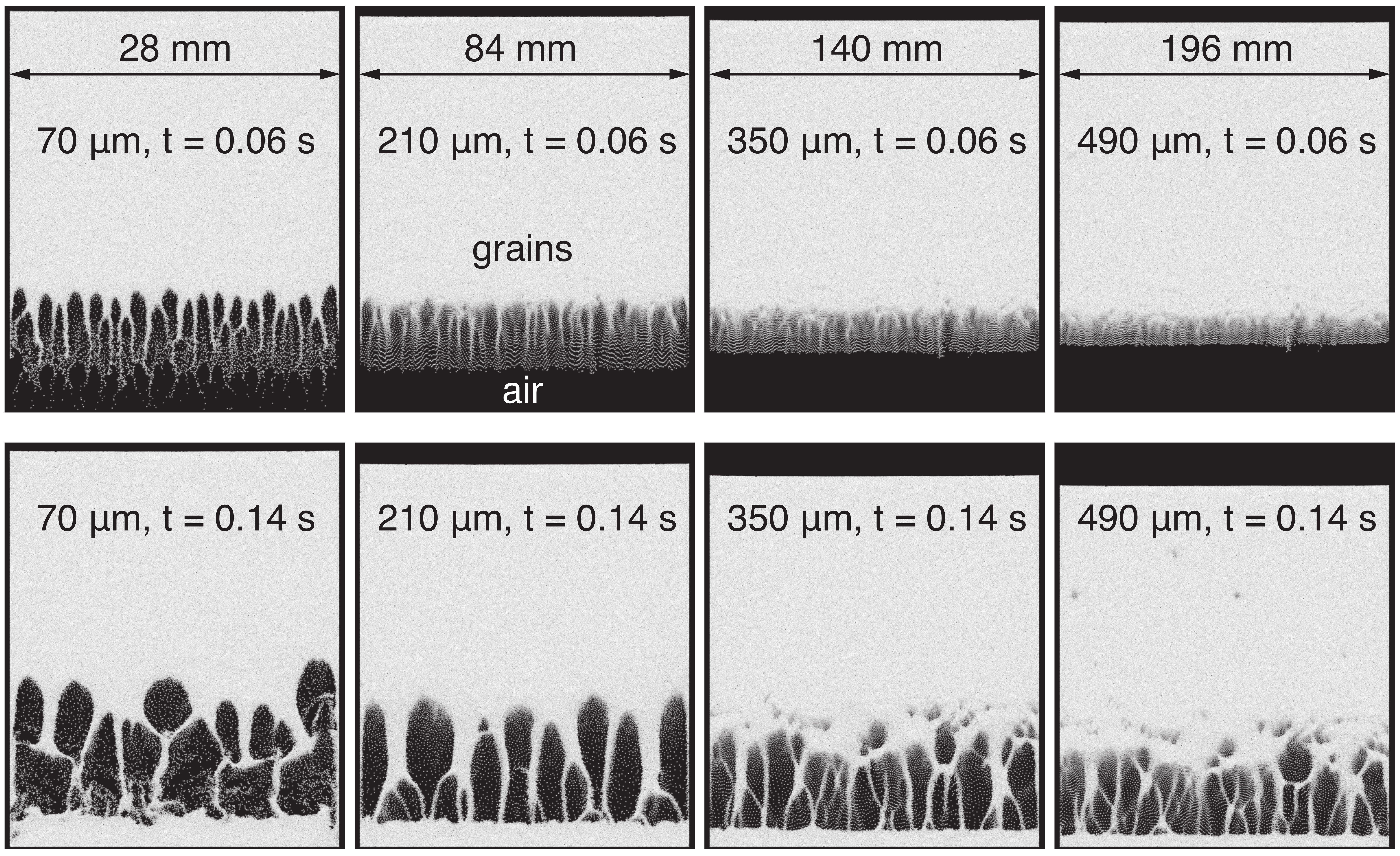} }
  \end{center}
  \caption{Simulation at four different grain/system sizes (horizontal
    axis) at two different times (vertical axis) showing the
    development of the instability with the initial front
    destabilization in the top row.}
  \label{GrainDiaDestab}       
\end{figure}

\begin{figure}
  \begin{center}
    \resizebox{0.95\columnwidth}{!}{%
      \includegraphics{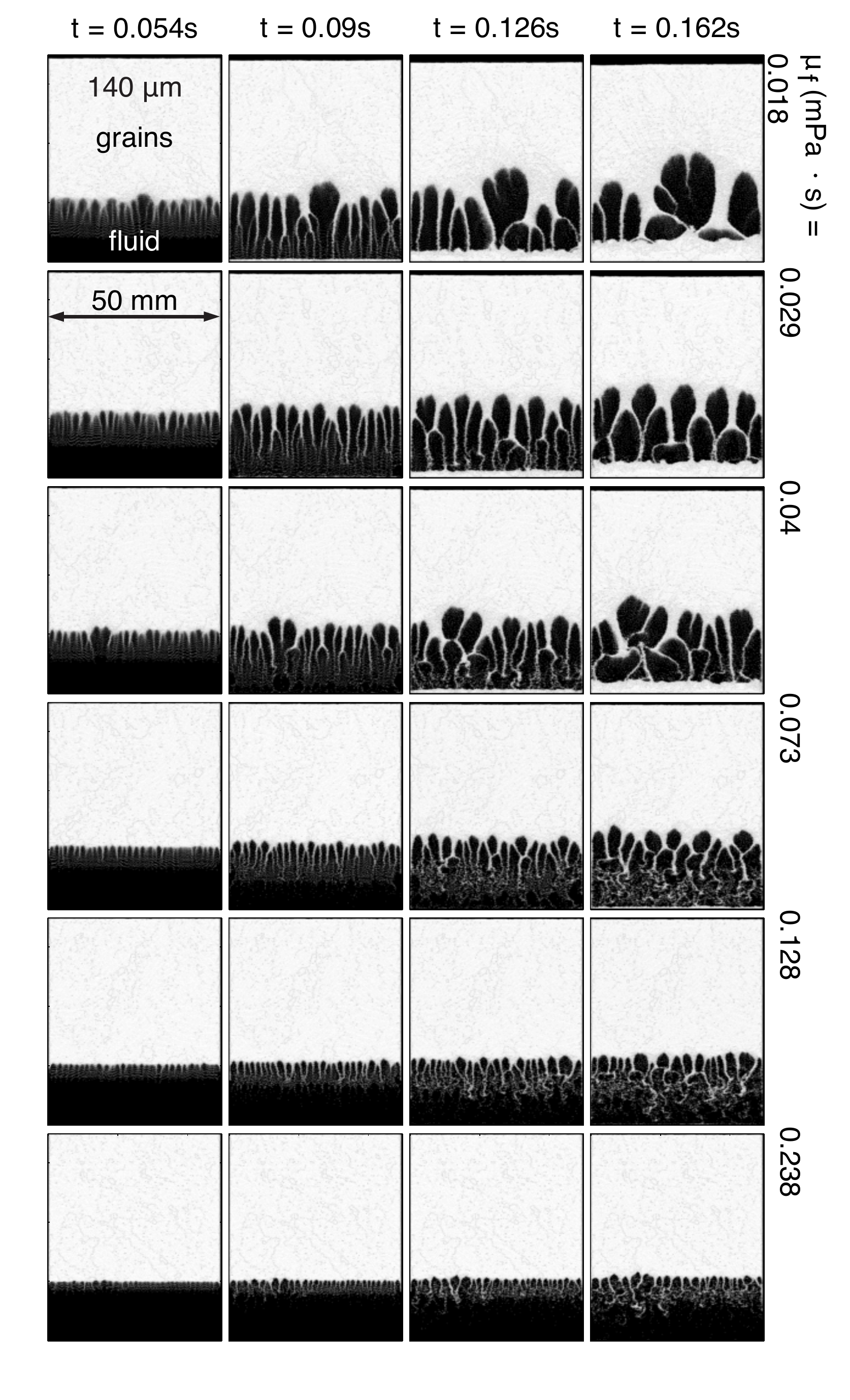} }
  \end{center}
  \caption{Numerical simulations of grains sedimenting in an incompressible fluid of
    variable viscosities $\mu_f$ (increasing from top to bottom). Time runs from
    left to right.}
  \label{incomp}       % Give a unique label
\end{figure}

\subsection{Dynamics of the granular Rayleigh-Taylor instability}

In both experiments and simulations, for grains falling in air or in
water/glycerol, the dynamics can be qualitatively described as
follows: From the initial configuration where grains are packed above
the clear fluid region, the dense grains start to flow downward,
i.e. sediment, while the fluid passes through the grains. This
situation leads to the formation of fingers of particles detaching
from the lower boundary of the upper compact pack. These fingers of
particles regroup, and bubbles of low particle density form in the
pack, rise and coalesce.  This dynamics is illustrated in
Fig.~\ref{ExperimentalSnapshots} (experimentally) and
Fig.~\ref{GrainDiaDestab} (numerically), for the air/grain case, and
in Fig.~\ref{incomp} for the incompressible fluid/grain case.

We notice that at low viscosities, simulations with incompressible
fluids are reminiscent of the dynamics obtained with compressible
air. As the viscosity is increased (from top to bottom in
Fig. \ref{incomp}), the fingers get mixed more rapidly as they detach
from the lower front of the compact pack.  The grains have here a size
of 140 microns in diameter, and a density of 2.5 g/cm$^3$, the fluid
has the density of water.

\begin{figure}
\begin{center}
\resizebox{0.5\columnwidth}{!}{%
  \includegraphics{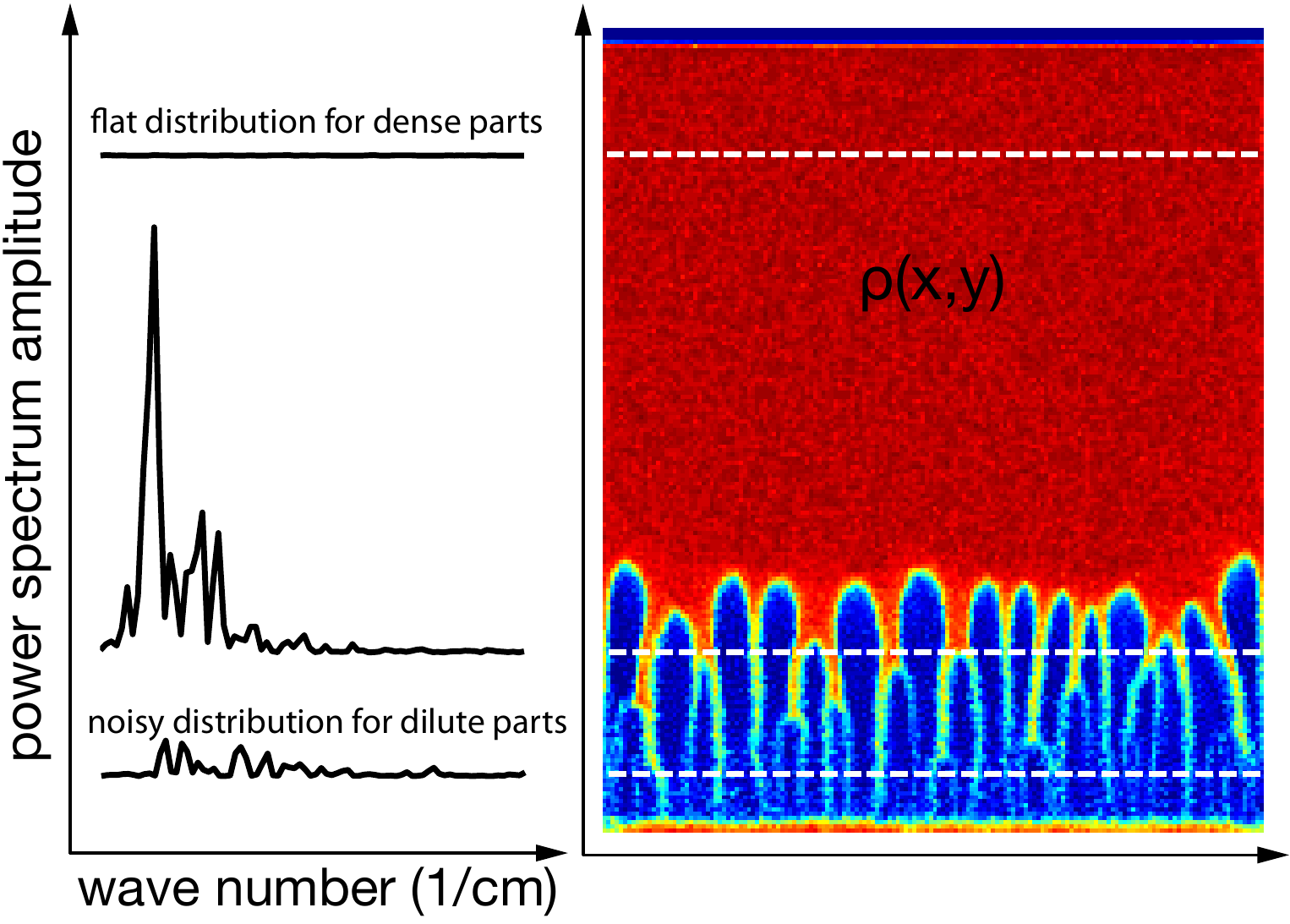} }
\end{center}
\caption{Extraction of a density function under the dense plug (middle
  dashed line in the right figure), and computation of its Fourier power spectrum
  (left).}
\label{Profile_extraction}       % Give a unique label
\end{figure}

%\subsubsection{Analyzing technique}
%\label{sec:analyze}
One way to analyze the instability of the detachment front, at the
bottom of the compact pack, is to compute horizontal density profiles.
In this way, it is possible to measure the lateral growth of density
fluctuations, while the fingers form and regroup. This can be done at
each horizontal level at vertical coordinate $y$ and time $t$,
extracting density fronts as function of the horizontal variable $x$,
$\rho(x,y,t)$. This is illustrated in Fig.~\ref{Profile_extraction}.

The most variable profiles as function of $x$ are just below the
detachment front: the particle density is roughly constant (at a scale
above a few grains) in the compact packing above the detachment front,
and constant and low in the empty region below. Hence, to increase the
statistical quality of the quantity we analyze, we perform the average
over all horizontal profiles at a given time, and obtain in this way
horizontal density profiles $\rho(x,t)=\langle\rho(x,y,t)\rangle_y$.
For the experiments, the density is not known directly, but the same
procedure is performed on the basis of light intensity (pixel gray
value) recorded with a fast camera.

\begin{figure}
\resizebox{\columnwidth}{!}{%
 \includegraphics{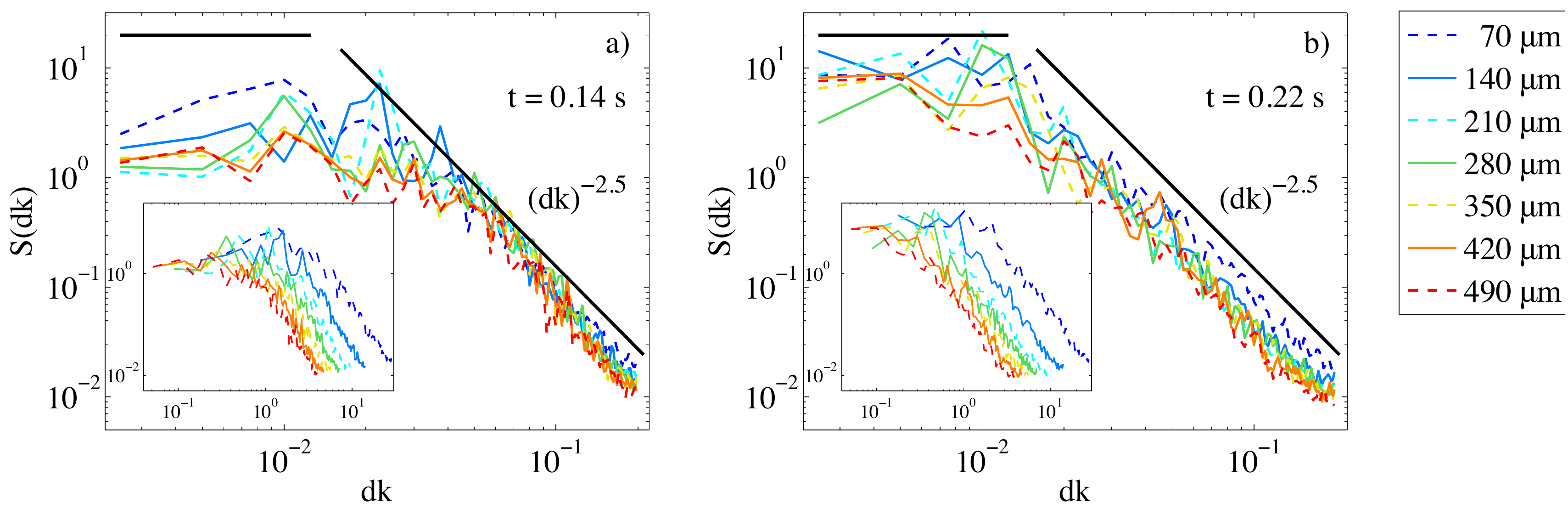}}
\caption{Power spectrum collapse, simulations: spatial scaling.}
\label{ScaleInSimu}%
\end{figure}

\begin{figure}
\resizebox{\columnwidth}{!}{%
 \includegraphics{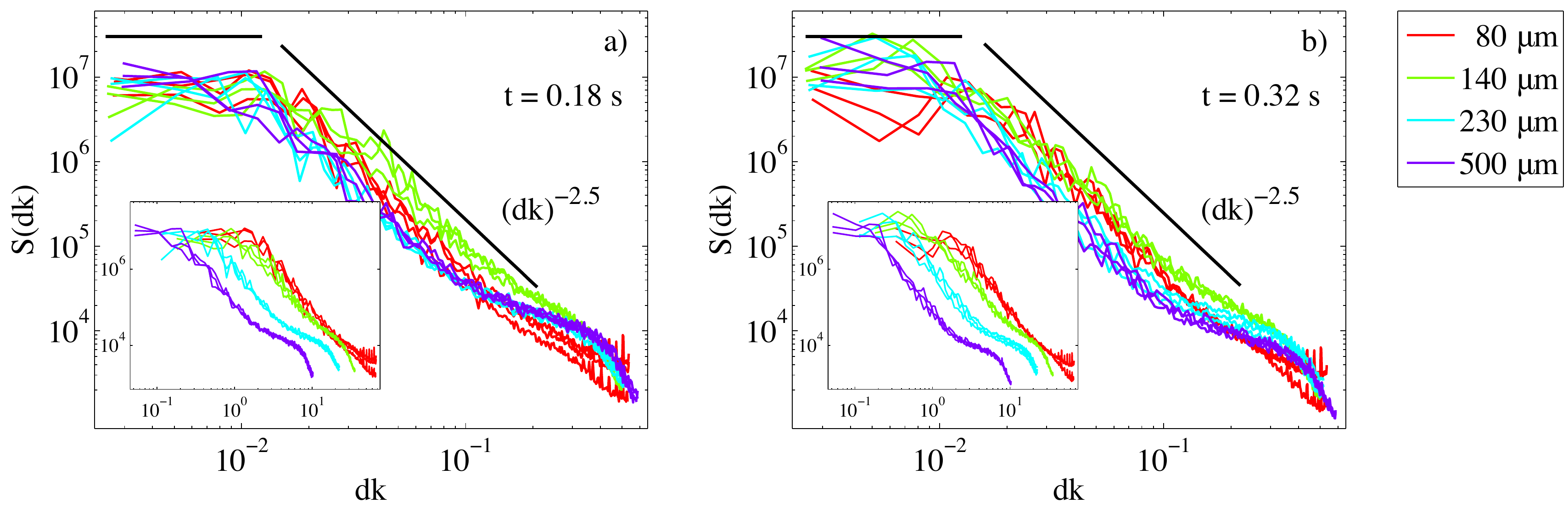}}
\caption{Power spectrum collapse, experiments: spatial scaling.}
\label{ScaleInExps}%
\end{figure}

These horizontal density functions are then spatially Fourier
transformed, and their Fourier power-spectra are represented at
various times and for various grain sizes using a bi-logarithmic
scale, as is shown in Fig.~\ref{ScaleInSimu} for the simulations, and
in Fig.~\ref{ScaleInExps} for the experiments. The inset plots
represent the raw power spectra for all sizes, and the full figures
indicate the power spectra using the grain diameters as a unit
size. The comparison between both shows the size collapse.
 
The experiments and simulations with compressible air have been
analyzed as function of the particle size.  Snapshots of the early
times of these simulations with variable particle sizes are shown in
Fig.~\ref{ExperimentalSnapshots}. A spatial rescaling using the grain
diameter rendered the pattern formation mechanism similar for the
range of diameters investigated. This was studied in details in
\cite{vinningland10}. The rescaling is illustrated by the collapse of
the structures analyzed in Fig.~\ref{ScaleInSimu} (for simulations)
and Fig.~\ref{ScaleInExps} (for experiments), obtained after rescaling
the spatial units with the grain diameter, $d$. The insets in these
figures show the density profiles prior to rescaling.

\section{Family-Vicsek scaling of the lower decompaction front}
In the plots of the Fourier transformed density profiles in
Figs.~\ref{ScaleInSimu} and \ref{ScaleInExps}, one notes that, both
for the experiments and for the simulations, the low wavenumber part
of the spectra is flat. This corresponds to spatially uncorrelated
density fluctuations for large scales (with a white noise character),
whereas the power spectrum decreases as a power-law for small scales
(high wavenumber). A behavior $$S(k,t) \sim k^{-2.5} $$ is observed
on this part, both for experiments and for simulations. The cutoff
$k_c$ corresponds to the maximum size of the density fluctuations,
$\xi=2\pi/k_c$, called the correlation length. The power-law behavior
for scales smaller than the correlation length corresponds to the
self-affine character of these density fluctuations. The power
spectrum distribution of a self-affine quantity may be expressed as
$$S(k,t) \sim k^{-(1+2\zeta)},$$ which in our case yields a Hurst
exponent $\zeta \sim 0.75$. 

Deviations are observed for the experiments in Fig.~\ref{ScaleInExps}
on the smallest scales, which can be attributed to a different
dynamics at scales of the order of the plate separation.

\begin{figure}
\begin{center}
\resizebox{\columnwidth}{!}{%
  \includegraphics{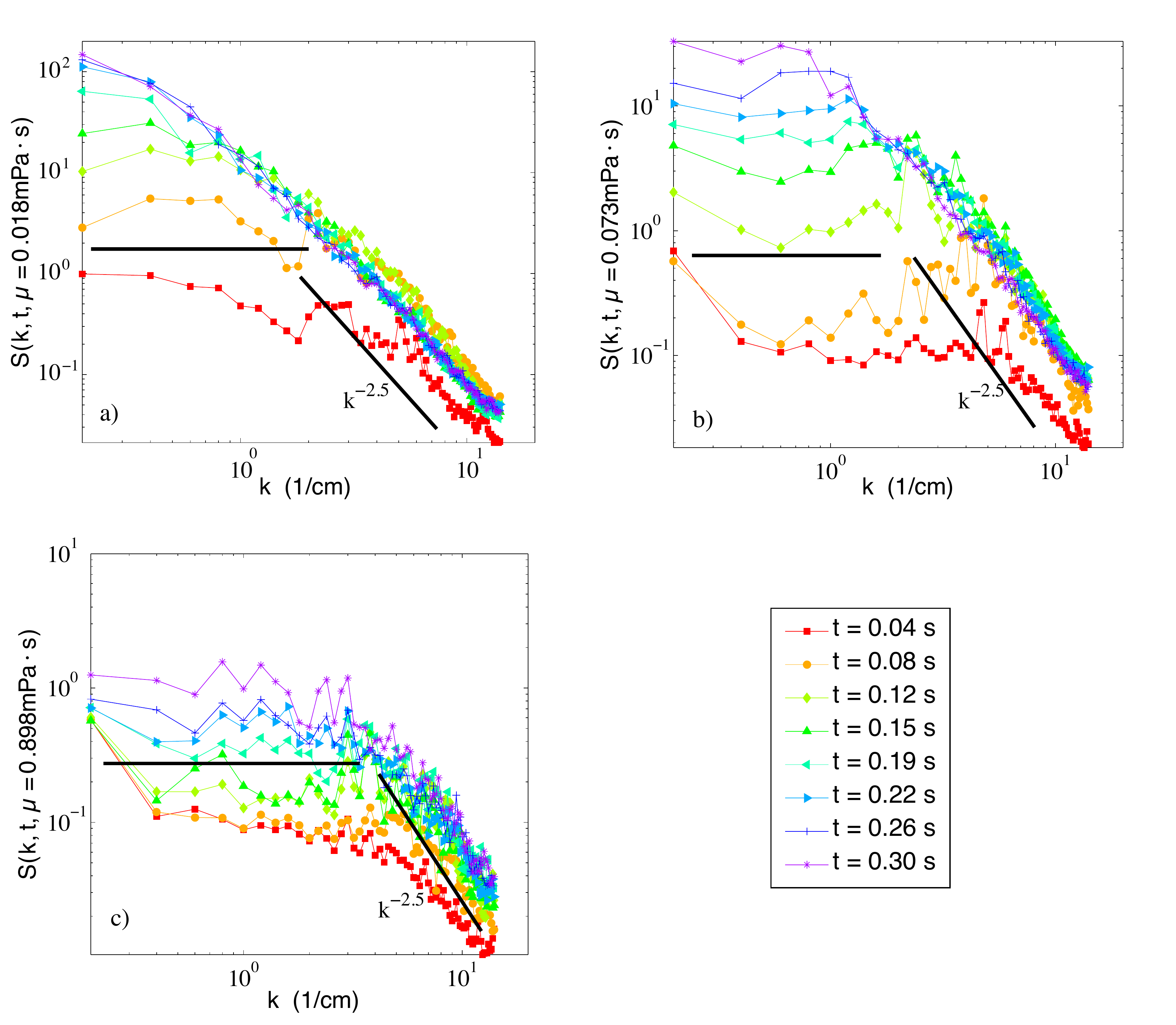} }
\end{center}
\caption{Power spectra of the horizontal density profile for different
  times: results of simulations in an incompressible fluid. The
  viscosities of the fluids are respectively (a): 0.018, (b): 0.073
  and (c): 0.9 mPa$\cdot$s. The different curves correspond to
  different times after the start of the instability.}
\label{Scale_incomp} 
\end{figure}

Interestingly, when the case of incompressible fluids is compared, a
similar behavior is observed as for the compressible case, see
Fig.~\ref{Scale_incomp}. This behavior extends to the whole range of
viscosities probed, from 0.018 to 0.9 mPa$\cdot$s, almost three orders of
magnitude. This is the case despite the obvious difference between the
developed stages of the instability, as is seen by comparing
Figs.~\ref{GrainDiaDestab} and \ref{incomp}. The early stages,
however, seem to follow the same type of behavior with respect to the
lateral spreading of the density fluctuations.

\begin{figure}
  \begin{center}
    \resizebox{.9\columnwidth}{!}{%
      \includegraphics{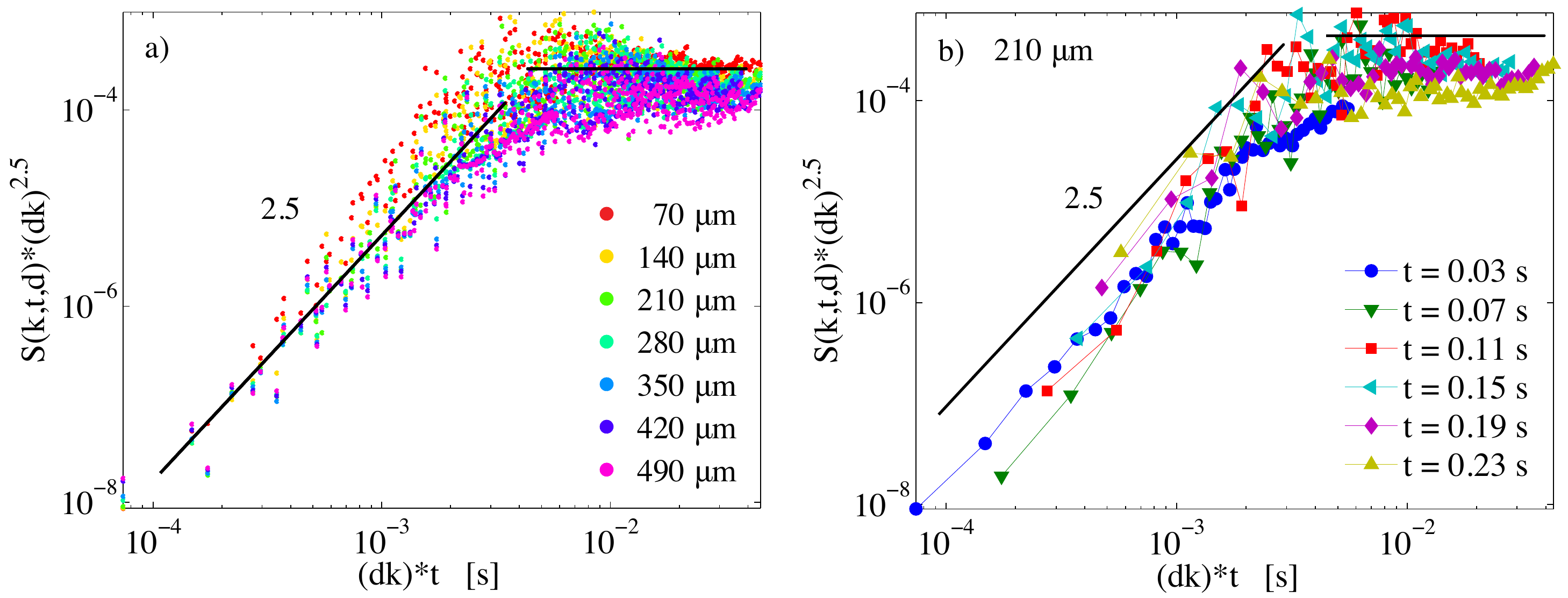} }
  \end{center}
  \caption{Family-Vicsek scaling obtained from numerical data. (a)
    shows the scaling for all grain sizes and for $t \le$ 0.23 s,
    while (b) is the same plot as (a) but only showing the 210 \um
    data and where the colors indicate the time from blue (early
    times) to red (late times).}
  \label{FVcomp}
\end{figure}

\begin{figure}
  \begin{center}
    \resizebox{.9\columnwidth}{!}{%
      \includegraphics{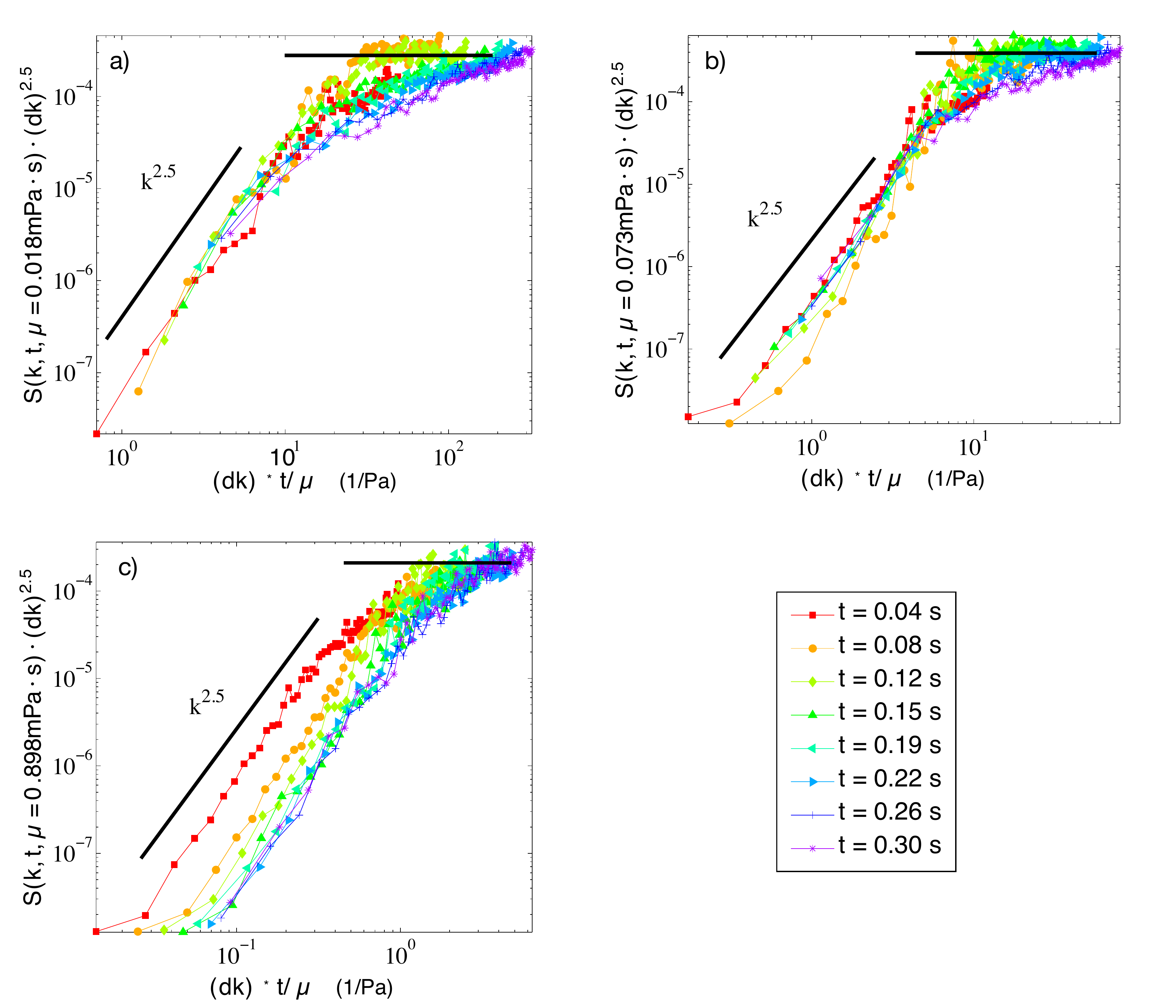} }
  \end{center}
  \caption{Master function intervening in the Family-Vicsek scaling for
    the Rayleigh-Taylor instability in incompressible fluids: Scaled
    power spectra $S(k,t) * (d*k)^{2.5} $, which are functions $g$ of
    $d*k*t/\tau$, with a behavior $g(x) \sim C_1 x^{2.5}$ at small
    arguments, and $g(x) \sim C_2$ at large ones. The Hurst exponent
    used for the scaling is 0.75, and the dynamic exponent is 1. The
    viscosities of the fluids are respectively (a): 0.018, (b): 0.073
    and (c): 0.9 mPa$\cdot$s.}
  \label{FVincomp}
\end{figure}

The Family-Vicsek scaling behavior is summarized in the plots in
Fig.~\ref{FVcomp} for the compressible case, and in
Fig.~\ref{FVincomp} for the incompressible case. For the compressible
case, the transformed density functions $S(k,t,d)$ for all the probed
grain sizes $d$ and for times $t \le$ 0.23 s all fall on a single
master curve shown in Fig.~\ref{FVcomp}(a). For the incompressible
case there is, however, a deviation from the Family-Vicsek scaling for
the early times and largest viscosities probed, as seen in the
incomplete collapse for the first times in Fig.~\ref{FVincomp}(c).

The observed Family-Vicsek scaling merely reflects the following about
the power spectrum $S(k,t)$:
\begin{eqnarray}
S(k,t) &=& C k^{-1-2\zeta} g(d k (t/\tau)^{1/z}) \\
g(x)  &\sim& C_1 x^{1+2 \zeta}\quad\mathrm{for}\ x \ll 1
\mathrm{,}\ \mathrm{i.e. }\ 1/k \gg \xi\ \mathrm{with}\ \xi \sim t^{1/z}  \\
g(x) &=& C_2\qquad\quad\ \,\mathrm{for}\ x \gg 1 \mathrm{,}\
\mathrm{i.e.}\ 1/k \ll \xi\,,
\end{eqnarray}
where $\tau$ is a time-constant and $C$,$C_1$ and $C_2$ are constants.
So that $S(k,t)$ is independent of $k$ for scales above the
correlation length $\xi$, and independent of $t$ when the mode $k$ has
saturated, with a wavelength $1/k$ exceeding the correlation length.
This leads to a generalized diffusion behavior
\begin{equation}
S(k,t)\sim t^{(1+2\zeta)/z}
\end{equation}
when $1/k>\xi$, and
\begin{equation}
S(k,t)\sim k^{-(1+2\zeta)}
\end{equation}
when $1/k<\xi$.

This corresponds, in real space, to a behavior for the root mean
square of the solid density $$\sigma_\rho = \langle\rho^2 -
\langle\rho\rangle^2\rangle^{1/2}$$ following a Family-Vicsek scaling that can be
obtained form Parseval's theorem, as
\begin{eqnarray}
  \lefteqn{\sigma^2(L,t) =  (1/L) \int_{k=2\pi/L}^\infty S(k,t) dk} \\
  & = & (1/L) \int_{k=2\pi/L}^\infty C k^{-1-2\zeta} g(d k (t/\tau)^{1/z}) dk \\
  & = & (t/\tau)^{(2\zeta)/z}  (1/L) \int_{2\pi(t/\tau)^{1/z}/L}^\infty k' g(k') dk'\,,
\end{eqnarray}
i.e.
$$\sigma(L,t) =  (t/\tau)^{\beta} F(L/t^{1/z})$$
with
$$ \beta = \zeta/z\,. $$
Hence, the density profile under the detachment front saturates as a
self-affine function \cite{Barabasi} with a Hurst exponent $\zeta \simeq 
0.75$ for scales below $\xi$, and displays a superdiffusive
behavior with an exponent $\beta \simeq 0.75$. The root mean square
of the solid density, $\sigma_\rho = \langle\rho^2 - \langle\rho\rangle^2\rangle^{1/2}$
displays a behavior corresponding to
$$ \sigma_\rho \sim t^\beta.$$
Conversely, for scales $l$ above $\xi$, 
one has a self-affine behavior, 
$$\langle(\rho(x+l) - \rho(x)\rangle^2 \sim l^{2\zeta}\,.$$
The correlation length $\xi$ is increasing as a power law with time,
as $\xi \sim t^{1/z}$. The density in the detachment front exhibits
these behaviors with respectively dynamic, Hurst and growth exponents
\begin{eqnarray}
z & = & 1 \\
\zeta & = & 0.75 \\
\beta & = & \zeta / z = 0.75
\end{eqnarray}
One can also equivalently express the Family-Vicsek scaling with a
scaling function $g$, such as:
\begin{eqnarray}
S(k,t) & = & C (d k)^{-2.5} g(d k (t/\tau)) \\
g(x)  & \sim &  C_1 x^{2.5}\quad\mathrm{for}\,x \ll
1,\,\,\mathrm{i.e.}\ 1/k \gg \xi\ \mathrm{with}\ \xi \sim t \\
g(x)  & = & C_2 \qquad\ \ \mathrm{for}\,x \gg 1,\,\,\mathrm{i.e.}\ 1/k \ll \xi\,.
\end{eqnarray}
Indeed, $S(k,t) * (d*k)^{2.5} = C g (d*k*t/\tau)$ is represented for
air/grain simulations in Fig.~\ref{FVcomp}, and this scaling function
master curve displays these two expected behaviors for the
granular-gas flow.

This is also the case for the granular-incompressible fluid flows, as
shown in Fig.~\ref{FVincomp} for the three viscosities. As observed
in the previous representation, the collapse is rather followed, apart
from the earliest times at larger viscosities.

\section{Conclusion}
We have studied situations where well packed cohesionless grain
assemblies, heavier than the fluid between them, are released in a
clear fluid. This gives rise to the so-called granular Rayleigh-Taylor
instability, where bubbles empty of grains rise, while granular
fingers form in the clear fluid region.

The principles of hybrid simulations starting from basic physical
equations was established, for both cases where the carrier fluid is a
perfect gas or an incompressible viscous fluid. The experimental setup
and results are also shown to be consistent with the simulations in the
air/grain case.
 
Grains are mixed during this process, and one can follow the evolution
of spatial correlations in lateral density fluctuations, with
vertically averaged horizontal grain density profiles $\rho(x)$.
These density fluctuations are shown to follow an anomalous diffusion
behavior with a growth exponent $\beta\sim 0.75$. Hence, for two
points at a large horizontal distance $x$ from each other, a
difference in density will grow as $\delta \rho (x,t) \sim
t^\beta$. The density fluctuations will follow this behavior up to a
saturation when the correlation length $\xi$ reaches x. $\xi$ is
growing as a power-law of time, with a dynamic exponent around
1. Eventually, for scales larger than $\xi$, the density profile
displays a self-affine behavior, with a Hurst exponent around 0.75.
This is seen in the spatial Fourier domain, both for experiments and
simulations for air-grain systems, and for simulations for mixtures of
grains and incompressible viscous fluid.

Eventually, we have shown that the power-spectra of the density at all
times can be collapsed according to a Family-Vicsek scaling.

\end{document}